\documentclass[journal=ancac3,manuscript=article]{achemso} 

\usepackage{graphicx}
\usepackage{amssymb,amsmath}
\usepackage{color}
\usepackage{calc}
\usepackage{xspace}
\usepackage{mhchem}
\usepackage{soul} 

\author{Irina B. Tsvetkova}
\author{Arathi Anil Sushma}
\affiliation[Department of Chemistry]
{Indiana University, Bloomington, IN 47405, U.S.}

\author{Joseph C.-Y. Wang}
\affiliation[Department of Molecular and Cellular Biochemistry]
{Indiana University, Bloomington, IN 47405, U.S.}

\author{William L. Schaich}
\affiliation[Physics Department]
{Indiana University, Bloomington, IN 47405, U.S.}

\author{Bogdan Dragnea}
\affiliation[Department of Chemistry]
{Indiana University, Bloomington, IN 47405, U.S.}
\email{dragnea@indiana.edu}
\phone{+1 (812 8560087}

\title[]
  {Radiation Brightening from Virus-like Particles}

\abbreviations{VLP}

\begin{document}


\begin{abstract}
Concentration quenching is a well-known challenge in many fluorescence imaging applications. Here we show that the optical emission from hundreds of chromophores confined onto the surface of a virus particle 28 nm diameter can be recovered under pulsed irradiation. We have found that, as one increases the number of chromophores tightly-bound to the virus surface, fluorescence quenching ensues at first, but when the number of chromophores per particle is nearing the maximum number of surface sites allowable, a sudden brightening of the emitted light and a shortening of the excited state lifetime are observed. This radiation brightening occurs only under short pulse excitation; steady-state excitation is characterized by conventional concentration quenching for any number of chromophores per particle. The observed suppression of fluorescence quenching is consistent with efficient, collective radiation at room temperature. Interestingly, radiation brightening disappears when the emittersÕ spatial and/or dynamic heterogeneity is increased, suggesting that the template structural properties may play a role and opening a way towards novel, virus-enabled imaging vectors that have qualitatively different optical properties than state-of-the-art biophotonic agents.
\end{abstract}

\section{}

Photoluminescence, particularly fluorescence, is used in a myriad of applications in which low-background, high spatial resolution, and  rapid response are required for non-intrusive imaging, remote sensing, and control of transient chemical states. Thus, synthetic fluorophores are incorporated in sensors and detectors, e.g. for early warning of bio-aerosol threats,  used in operation rooms for intraoperative guidance in brain and prostate cancer surgery\cite{Tipirneni2017}, and in anti--counterfeiting materials\cite{Song2018}.  Over the years spectacular improvements in chromophore photostability, wavelength range, biological integration, detectors and detection techniques, have pushed the limits of fluorescence imaging to realms never thought possible before\cite{Stepanenko2011, Tsien2008}. Despite these improvements, and somewhat surprisingly in the context of ever more demanding cutting edge applications, fluorescence emission is still overwhelmingly by way of uncorrelated, random emission from multiple chromophores. Associated with this regime are undesirable characteristics such as self-quenching, when emitters are too close, and exponential decay that is relatively slow at molecular scale (typically, 1-5 ns). Extended excited state lifetimes limit emission brightness, increasing the likelihood of photobleaching,  and making the quantum yield more prone to change in response to environmental fluctuations\cite{Valeur2013}.

At the same time, a system's collective behavior can be much more than the sum of its parts and the optical characteristics of atomic systems are no exception\cite{Khitrova2007a}. Specifically, indistinguishability of microscopic quantum constituents can lead to coherent behavior via quantum, symmetry-enforced selection rules\cite{Dicke1954}. Such collective relaxation processes can be much faster and more efficient than those encountered near the thermodynamic limit. Indeed, certain molecular phenomena that are central to life on Earth appear to rely on quantum coherence\cite{Ball2011}. Thus, light-harvesting complexes - macromolecular assemblies in photosynthetic organisms, exploit the tight spatial organization of proteinaceous multi-chromophore assemblies to enable coherent exciton transport in short times and over distances many times larger than the size of molecular constituents\cite{Scholes2011}.  At least in one case\cite{Monshouwer1997}, it appears that the same molecular organization that is responsible for quantum coherence at room temperature in light harvesting complexes also leads to chromophore coupling and an accelerated radiative rate and suppression of fluorescence quenching. 

Control of the probability of light emission through cooperation among emitters has a venerable history\cite{Purcell1946}. It includes phenomena such as lasing, photon echo, magnetic resonance decay, amplified spontaneous emission (ASE), super-fluorescence (SF), and super-radiance (SR). A few of these phenomena have been already worked into powerful spectroscopy tools. 

In SR\cite{Dicke1954} (SR) and SF\cite{Vrehen1980} (SF) for instance, a sub-wavelength ensemble of indistinguishable two-level quantum emitters is prepared in (SR), or spontaneously attains a (SF), collective excited state which couples into a common electromagnetic field as a single, giant radiating dipole\cite{Gross1982a, Benedict1996, Cong2016}. Because of its substantially-accelerated radiative decay as the number of cooperating chromophores increases, in the years before the invention of the laser, SR was suggested as a way towards generating short coherent pulses of light, without a cavity. However, despite increased interest from biomedical applications requiring deep-subwavelength intense light sources\cite{MallawaArachchi2018,Wang2016}, this promise has proven difficult to fulfill. The reason is that coherent behavior emerges only when the quantum emitters are indistinguishable and the cooperative radiative decay of the system is faster than any other phase breaking (decoherence) process. Thus, for non-interacting (non-aggregated) emitters such as atomic gases\cite{Skribanowitz1973a}, molecular centers in crystals\cite{Malcuit1987}, and quantum dots\cite{Scheibner2007}, SR could be observed at low temperature only. On the other hand, strong coupling systems such as molecular crystals and arrays\cite{Spano1989} and including the light-harvesting complexes of certain photosynthetic microorganisms\cite{Monshouwer1997} exhibit accelerated radiative decays from excitonic states at close to room temperature. Only recently, coherent emission at room temperature was observed in ensembles of radiatively-coupled, but non-interacting in the quantum tunneling sense, diamond N-vacancies\cite{Bradac2017}.

Here we present experiments suggesting that coherent emission at room temperature from emitters coupled through their radiation field, but independent in terms of quantum tunneling coupling, can even occur in soft matter, namely, in single virus-like particles (VLPs) decorated with organic dyes, provided three conditions are met:
\begin{itemize}
\item First, the number of dyes per particle exceeds a critical value. Interestingly, at and above this critical number of dyes per particle, emission is nearly completely quenched under steady-state excitation. 
\item Second, pulsed irradiation is required for recovery of fluorescence emission, \emph{i.e.} for the radiative rate to overcome the non-radiative rate from concentration quenching. However pulse coherence does not seem to carry an important role. The effect is observable with either incoherent or coherent (transform-limited) pulses.  
\item Third, factors that induce static or dynamic fluctuations in either the template or the chromophore network, also suppress radiation brightening. Thus, the radiation brightening effect is conditioned by the interaction of the chromophores with an ordered molecular scaffold, of a size much smaller than the wavelength of light.
\end{itemize}

Several collective emission phenomena that could be responsible for the observed radiation brightening in VLPs are considered. Although the treatment is not exhaustive and further research concerning the mechanism is warranted, the spectroscopic and dynamic evidence provided here are consistent with super-fluorescence. Therefore, dye-decorated VLPs represent promising candidates for a molecular embodiment of the so-called "Dicke optical bomb"\cite{Khitrova2007a}. Moreover, the effect can be obtained at room temperature, from genetically modifiable viromimetic particles through mild reactions and self-assembly.  When fully optimized, such biophotonic particles may fill a technological void in biological applications requiring bright, deep-subwavelength light sources. 

\section{Results}

\begin{figure}[ht]
  \centering
 \includegraphics [width =  \textwidth] {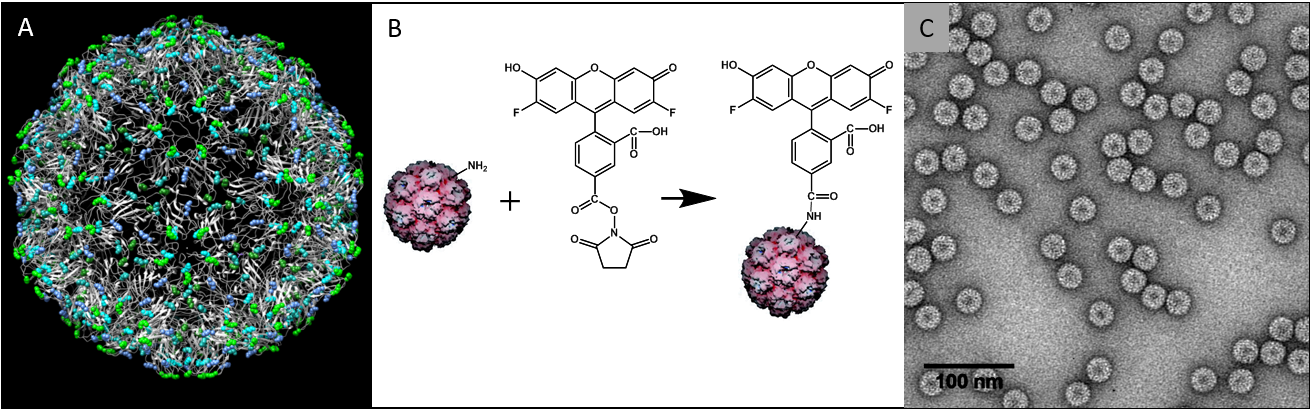}
  \caption{A) Molecular model of BMV with structure surface exposed lysines. B) OG-BMV conjugation reaction scheme. C) Transmission electron microscopy (TEM) image of near-saturation labeled BMV.}
  \label{f:constr}
\end{figure}

The molecular scaffold used for this work is a Brome mosaic virus (BMV) particle, which has an icosahedral capsid formed of 180 identical self-assembled proteins\cite{Lucas2002}. A common bioconjugation reaction of surface-exposed lysines was employed to covalently attach the fluorescein derivative dye, Oregon Green 488\texttrademark~carboxylic acid succinimidyl ester (OG), at specific locations on the virus, Fig.~\ref{f:constr}. The average number of chromophores was controlled by varying the molar ratio between capsid and dye in the reaction media, and by adjustment of buffer conditions (see Supporting Information -- Virus Modification section, for details on synthesis and characterization of the conjugation reaction results). 

There are a total of 12 lysines per BMV coat protein. However, matrix-assisted laser desorption ionization time of flight (MALDI-TOF) spectrometry indicated  that the maximum number of dyes per protein is 2. Interestingly, as the reaction evolves in time, the mass peak corresponding to a mass shift of 2 dyes appears after the mass shift for 1 dye/protein reached a maximum. This indicates that there are two accessible lysines, with different reactivity (see Fig. SI3).

\begin{figure}[ht]
\centering
\includegraphics[width = \textwidth]{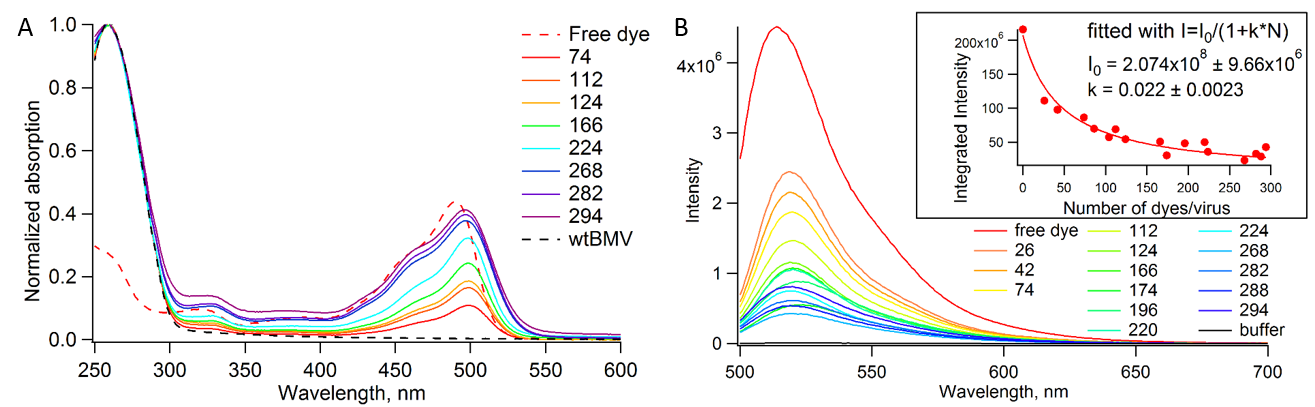}
\caption{Steady-state UV-Vis absorption (A) and fluorescence emission (B) spectra of free dye and BMV-OG. Inset in (B) shows fit corresponding to the Stern-Volmer equation which describes concentration quenching for spectrally-integrated, steady-state fluorescence intensity as a function of $<N>$.}
\label{f:cwspec}
\end{figure}

The degree of labeling was estimated from UV/Visible absorption spectra of purified BMV-chromophore conjugates (see Supporting Information -- Virus Modification and Characterization section and Fig.~\ref{f:cwspec}A), and confirmed by MALDI-TOF (Fig. SI3 in Supporting Information). 

The smallest distance between two nearest-neighbor lysines on the virus surface is $\sim19$ \AA~(see table in Fig. SI1). Therefore, unlike the case of J and H molecular aggregates\cite{Spano2010}, negligible nearest neighbor electronic coupling is expected here. Indeed, no changes in the form of spectral shifts or broadening/narrowing of the bands could be observed in the absorption and emission spectra as a function of $<N>$, at low excitation intensity, Fig.~\ref{f:cwspec}. Concentration quenching was, however, noticeable in steady-state fluorescence emission spectra, Fig.~\ref{f:cwspec}B. This is consistent with previous works on fluorescent VLPs\cite{Wen2016}, although, as reported for other dye-labeled VLPs\cite{Soto2006}, the quenching rate was below that typically observed in free-dye solutions.

Absorbance spectra showed a bathochromic shift for the main absorbance peak of dye-conjugated BMV samples, independent of $<N>$, from $\lambda = 491$ nm for the free dye, to $\lambda = 496$ nm, indicating that the dye is coupled to the protein, Fig.~\ref{f:cwspec}A.

To obtain further information on the dye-surface geometry, we have performed single particle reconstruction of cryo-electron microscopy images at $7~\mathrm{\AA}$ resolution. The structures of BMV and dye-saturated BMV are indistinguishable (see Fig. SI4). Interestingly, although no extra density could be attributed to the presence of OG molecules, the OG-conjugated BMV particles (BMV-OG) are significantly more stable to physical and chemical manipulations than wtBMV. Specifically, when dialyzed in disassembly buffer conditions for 24 hours, wtBMV is completely disassembled, while dye-saturated BMV remains intact, with no noticeable morphological changes (see Fig. SI5 and related discussion in the Virus Modification and Characterization section of the Supporting Information document). 

The added resilience of the BMV-OG particles to environmental challenges suggests that the chromophores interact strongly with the capsid, in a way that stabilizes it. Combined with the absence of detectable morphological changes at intermediate cryo-EM resolutions, we deduce that the chromophores should be tightly associated with capsid interfaces. Based on the increased capsid stability and structural indistinguishability from that of wtBMV, we posit that the supported chromophore network faithfully follows capsid structure.

We now turn our attention to the occurrence of radiation brightening by accelerated emission. To this end, we have first measured the effective fluorescence lifetime as estimated from the mean photon arrival times, and the time-integrated photon counts from single BMV-OG particles as a function of the average number of chromophores per particle, by fluorescence lifetime microscopy (FLIM) with time-correlated single photon counting (TCSPC) (see FLIM section in the Supporting Information document). 

The emission rate of an ensemble of randomly emitting independent radiators scales linearly with $<N>$. For a coherent or partially-coherent ensemble, however, the emission rate is expected to scale nonlinearly with $<N>$\cite{Allen1970}. Since the dephasing rate scales linearly with $<N>$\cite{Shammah2017a}, coherent behavior will be favored under strong population inversion and/or reduction of dephasing rates. To obtain a strong population inversion, for single particle investigations by FLIM we have used pulsed excitation from a supercontinuum laser (120 ps pulse duration, maximum $500~\mu$W average power at the sample). For bulk spectroscopy, we have used 170 fs transform-limited pulses from an optical parametric amplifier.

\begin{figure}[ht]
\centering
\includegraphics[width = \textwidth]{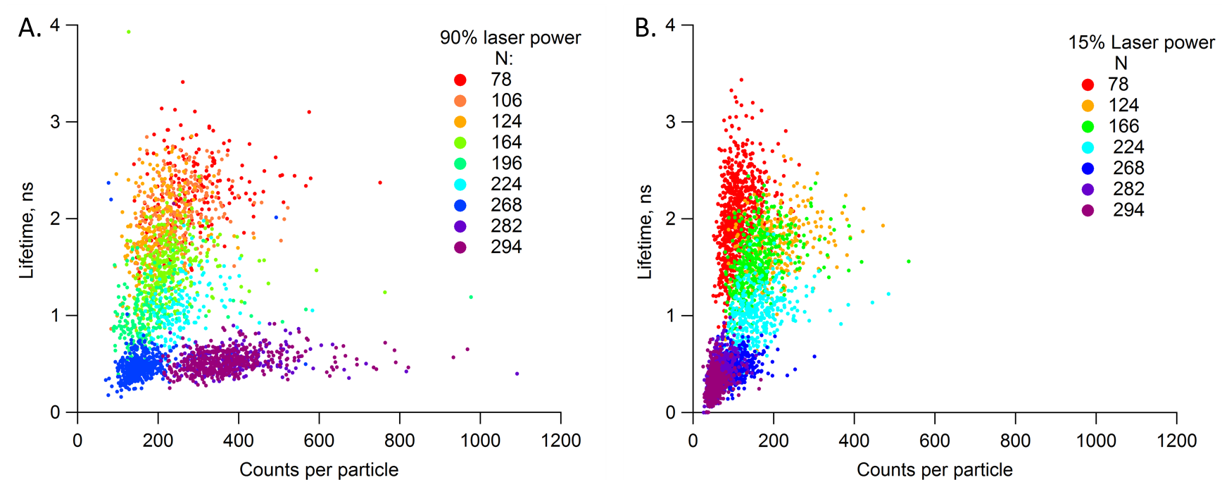}
\caption{A. Scatter plots of estimated lifetime \emph{vs.} fluorescence photon count collected from hundreds of labeled BMV particles by single-particle FLIM (each dot represents a single virus particle). Different colors represent samples with different average number of chromophores per particle, $<N>$). B. Same as A, except for the excitation power, which is $6\times$ less than in A. }
\label{f:IvsT}
\end{figure}

Fig.~\ref{f:IvsT} shows the distributions of the fluorescence lifetime and time-integrated photon counts from single OG-BMV particles with different $N$s under supercontinuum pulsed irradiation. In  Fig.~\ref{f:IvsT}A, FLIM was carried at 90\% of the maximum laser power (i.e. $\sim 450~\mu$W at the sample), while in Fig.~\ref{f:IvsT}B only 15\% of the maximum laser power ($\sim 75~\mu$W at the sample) was used. 

 In both, Fig.~\ref{f:IvsT}A and \ref{f:IvsT}B, as the number of chromophores per particle increases to $<N> \approx 200$, both photon counts and lifetime decrease. This is consistent with an increase in the nonradiative relaxation rate, as one may expect from the concentration quenching effect already observed by steady-state spectroscopy (see inset of Fig.~\ref{f:cwspec}). However, as N increases above $\sim 200$, bright particles with short lifetimes appear in Fig.~\ref{f:IvsT}A, but not in \ref{f:IvsT}B. This is consistent with a significant avoidance of concentration quenching. 
 
What collective emission phenomena may be responsible for the observed radiation brightening in VLPs? From the onset we can exclude lasing, \emph{i.e.} light amplification by stimulated emission, because of the very small VLP diameter, compared with the wavelength, and the absence of a resonator. Moreover, the density of chromophores in our case is well below that calculated for the stimulated emission threshold in nanolasers, for instance\cite{Shahbazyan2017}. Random lasing like that observed in nanoparticle suspensions\cite{Samuel2009} is an unlikely explanation, due to the fact that radiation brightening is witnessed at the scale of an individual particle. Furthermore, as we shall see later, we do not observe the narrow linewidth typical to lasing. 

Other possibilities include SF, SR, and ASE.  It is useful at this point to provide some background. While SF and SR are closely related,  it is generally considered that SR results when the collective polarization is generated by an external coherent laser field, while SF occurs when the atomic system is initially incoherent, and the collective polarization develops spontaneously from quantum fluctuations; the resulting macroscopic dipole decays superradiantly at the last stage\cite{Vrehen1980}. In both cases, emission is in the form of a pulse.  The SF (or SR) pulse duration, $\tau_R$, is inversely proportional to $N$, the number of coupled chromophores:
\begin{equation}
 \tau_R \sim \tau_{sp}/N. 
 \label{e:tauN}
 \end{equation}
 where $\tau_{sp}$ is the spontaneous fluorescence decay time for a single molecule.
 In the most basic form of SR, for a two-level system\cite{Dicke1954}, the peak intensity scales as the square of the number of radiators:
 \begin{equation}
 I_{max} \sim N^2.
 \end{equation}
 
In ASE\cite{Allen1973}, chromophores absorb light, and start decaying by spontaneous emission. If the emitted light passes through a region containing inverted molecules, it may de-excite these, being amplified in the process. The decay accelerates, shortening the emitted pulse duration and narrowing the spectral profile of the system. Thus, there are qualitative differences between ASE and SF emission in both time and frequency domain. In the frequency domain, ASE occurs in a narrower spectral window than the emission of an uncorrelated ensemble. SF(SR) emission occurs in a broader window than that of a single radiator\cite{Dicke1954}.  In the time domain, the SF pulse is sharp and develops after a time delay, which is typically  $\sim 10-50 \times \tau_R$\cite{Malcuit1987}. In ASE, the time delay is vanishingly small, the output pulse is longer and very noisy. Important for our case, ASE pulse duration is not sensitive to dephasing processes.

\begin{figure}[ht]
\centering
\includegraphics[width = \textwidth]{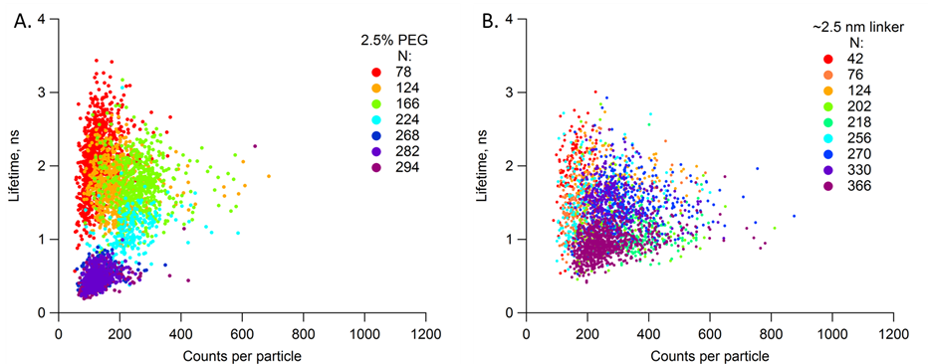}
\caption{Scatter plots of estimated lifetime vs fluorescence photon counts at same laser power as in Fig.~\ref{f:IvsT}A but in the presence of 2.5\% PEG 6000 (A), and with flexible linkages between the surface lysines and the chromophores (B). }
\label{f:ctrl}
\end{figure}

Since ASE is not sensitive to phase breaking processes, one would expect ASE to occur even in the presence of static and dynamic fluctuations (disorder), provided the intensity is above a threshold value. To test this idea, we added polyethylene glycol (PEG, MW 6000) to the dye-BMV solution. PEG exerts depletion forces on the viral coat protein, and can significantly increase the osmotic pressure on the virus shell\cite{Gelbart2009}. As a result, in BMV, PEG osmotic pressure induces a measurable ($\sim 10$ \AA) average radius compression at concentrations above 2.5\%\cite{Zeng2017b}. Furthermore, molecular interactions between PEG and the virus surface are likely to disrupt the protein/dye interaction and/or change the local environment of the dye\cite{Arakawa1985, Knowles2015a}. Thus, the distribution of nearest-neighbor distances and of orientations is likely perturbed in this case. Upon 2.5\% PEG addition, the brightening effect observed in unperturbed VLPs is suppressed, even at the highest laser intensity, Fig.~\ref{f:ctrl}.

In an additional experiment investigating the role of dye-virus template interaction, we have added a 25 \AA~linker between the chromophore and the coat protein (see reaction scheme in the Supporting Information document, Fig. SI9),  which decreased the possibility of chromophores interacting directly with the virus shell, and pushed the chromophore into solution, thus making them more prone to positional fluctuations, Fig.~\ref{f:ctrl}B. Adding the linker has not changed the absorption spectrum, which suggests that the chromophores remain separated. 

Similar to the PEG experiment, adding the linker significantly disrupted the brightening trend in Fig.~\ref{f:IvsT}A, and overall increased the data spread. Note that, under this approach, some of the dye molecules were still conjugated directly to the surface-exposed lysines, which further added to the heterogeneity of the dye environment (see Supporting Information -- Additional Supporting Figures for Control Sample Characterization).  Increased multiplicity of sites is a likely explanation the observed increase in the spread of measured brightness values, Fig.~\ref{f:ctrl}B. The overall trend is consistent with a decrease of intensity and lifetime as one would expect from concentration quenching. Therefore, it appears that a variety of disruption modalities of the chromophore-template arrangement leads to the same result of cancellation of the radiation brightening effect.

To further investigate the correlation between the template ordering and occurrence of the brightening effect, in the next experiment we have used a glass bead, $\sim27$ nm diameter, instead of the virus template (see Supporting Information -- Silica Nanoparticles Control Sample). The beads were functionalized with amines so that similar chemistry for dye attachment as for VLPs could be used. Life-time/brightness results and steady-state spectra for dye-coated silica glass particles are presented in Fig. SI10. There is increased heterogeneity in all data, the absorbance spectra  showing significant heterogeneous broadening. Avoidance of fluorescence self-quenching is not observed, at least within the pump power range available to us.  

\begin{figure}[ht]
\centering
\includegraphics[width = 0.9 \textwidth]{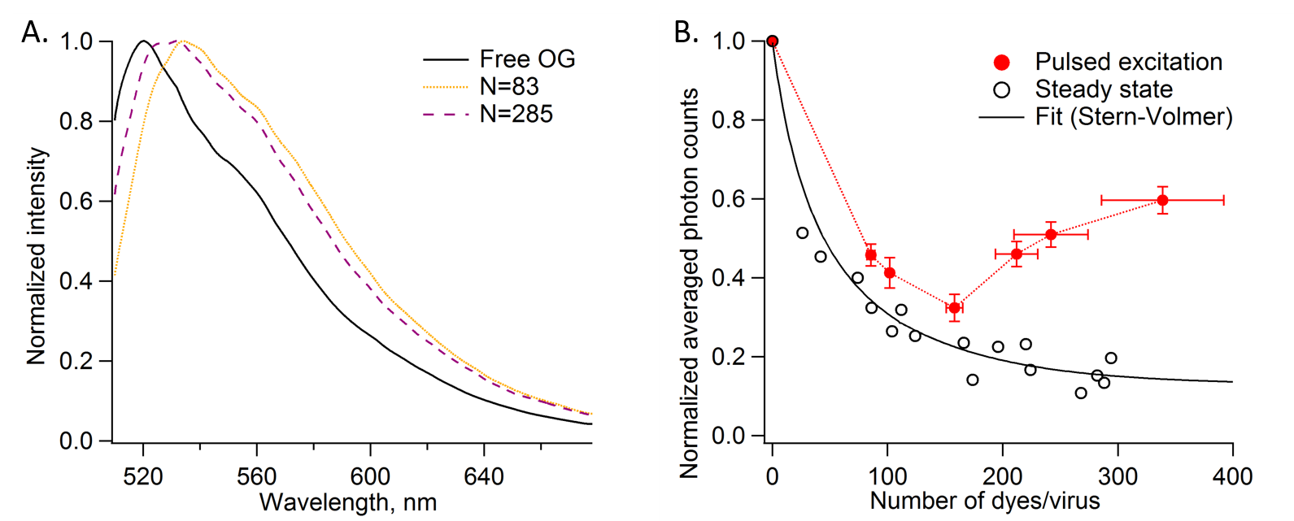}
\caption{A. Normalized fluorescence spectra for free dye and samples with ÒlowÓ and ÒhighÓ $<N>$ obtained under pulsed excitation at population inversion conditions, B. Normalized integrated photon counts \emph{vs.} $<N>$ for steady state, and pulsed excitation conditions. }
\label{f:OPA}
\end{figure}

Returning to the question of the expected emission linewidth difference between ASE and SF/SR, we collected ensemble fluorescence spectra under pulsed, focused excitation. In this case, unlike in FLIM experiments, we have utilized an OPA source that delivered transform-limited ultrafast pulses. Under population inversion conditions, we observed very similar spectral characteristics as for steady-state conditions, Fig.~\ref{f:OPA}A, except for the relative brightness, which increased. There is no evidence of spectral narrowing of emission. The quantitative analysis of the integrated photon counts vs $<N>$, Fig.~\ref{f:OPA}B, clearly shows an increase at dye high coverages, after an initial decrease from low to mid coverages. The threshold where the take-off is observed is $<N> \approx 170$, \emph{i.e.} somewhat lower than that obtained in FLIM experiments. This lower threshold for radiation brightening could be the result of initial condition preparation, due to a difference in laser pulse coherence and duration, or it could be due to the fact that, in FLIM VLPs are adsorbed on a surface, which presumably locally perturbs the template/chromophore network arrangement.

To summarize the main results, we have found that: 
\begin{enumerate}
\item Radiation brightening occurs from VLP-supported dyes under pulsed irradiation when the number of dyes per particle exceeds $\sim170-200$. Above this threshold, there is radiation enhacement from single VLPs with respect to the same particles at pump conditions where radiation brightening is not observed, Fig. SI8. 
\item Radiation brightening is anticorrelated with the presence of factors that affect the structure of the template and the chromophore-template interaction in ways that are known to increase disorder.
\item The VLP-supported dyes are coupled radiatively, but isolated from the point of view of stronger, direct interactions.
\label{S}
\item The radiation brightening effect correlates with population inversion.
\item There is no evidence for emission linewidth narrowing, like in nanolasing or ASE. 
\end{enumerate}

Item (\ref{S}) is intriguing: How could template order be involved? We hypothesize the virus template may play several roles: First, its organized structure ensures that the nearest neighbor distance between chromophores is always the same, across the entire chromophore set. As a consequence, the chromophores are weakly coupled through their radiation field, but otherwise isolated in the sense that there is no quantum tunneling. The latter makes the physics qualitatively different from that of the previously much more studied strong coupling in molecular aggregates and crystals \cite{Spano1989}. Moreover, coherent behavior is, in general, in competition with dipole-dipole dephasing due to Van der Waals interaction\cite{Gross1982a}. In a sample formed of many randomly-placed radiators, the dipole-dipole Van der Waals coupling is in general non-invariant by atom permutation because various radiators have different close-neighbor environments. It has been shown by several authors reviewed in ref.~\cite{Gross1982a} that these fluctuations are analogous to an inhomogeneous dephasing of dipoles, leading to correlation loss. It follows that, in a sample where the dipole-dipole coupling is invariant to permutation, as in an ordered array supported by a high-symmetry template, related decoherence mechanisms are reduced. 

Second, while in simpler coherent relaxation cases such as the Dicke two-level system SR, the geometric arrangement is inconsequential, in multiple-level systems template symmetry  can play a role in excited state degeneracy; Fano resonances in such systems are expected to strongly influence the emission rate from these levels\cite{Svidzinsky2008}. Thus, partly-coherent excitation pulses, such as those provided by the supercontinuum source, may set the VLP in an initial excited state from which the wavefunction evolves to a state that can rapidly cascade down to the ground level, and generate SF.

Third, strong association between the chromophores and the template is likely to further reduce the dynamic heterogeneity of the chromophore's local environment\cite{Noriega2015}, thus further mitigating the effects of dephasing processes. Moreover, since the virus shell is a nearly perfect spherical crystal, it is not unreasonable to consider the possibility of involvement of polariton coupling\cite{Dietrich2016}, with the polariton band structure being dictated by symmetry.

At this point, the mechanism for template involvement remains unclear. Further mechanistic investigations are warranted.  Fluorescence dynamics experiments with better temporal resolution could test for the existence of a self-organization time, which is a characteristic of SF.  Moreover, since the probability of two simultaneous photon emissions is a measure of the number of independently emitting regions\cite{Paul1982}, examining photon statistics through the measurement of the 2nd order correlation function will provide an estimate of the size of correlation region in a VLP and inform on how it depends on template order, $N$, etc. At this point, it is unknown whether, at intermediate chromophore loading levels, the chromophores are randomly distributed or they tend to form continuous network domains separated by large vacancies. Presumably, chromophore arrangement is characterized by significant fluctuations from one particle to another at intermediate loading levels. Future experiments could take advantage of the VLP approach significant level of structural control.  on the symmetry itself, with virus coat proteins being able to self-assemble in a variety of symmetric cages (e.g. T=1 with 60 proteins, pseudo T=2 with 120 proteins, and T=3 with 180 proteins) depending on assembly conditions\cite{Sun2007}. Furthermore, one unique advantage of the protein cage approach is the possibility of genetic encoding of photonic function and stoichiometric expression of fluorescent proteins\cite{Brillault2017} or incorporation of unnatural aminoacids\cite{Lang2014}. Such experiments are poised to answer many of the mechanistic questions opened by the observations reported here. The ability to avoid quenching in imaging or photonic control applications, especially in biomedical settings where biocompatibility is a paramount concern, makes radiation brightening in VLPs worth study.

In conclusion, we have observed radiation brightening via collective relaxation in solution, at room temperature, from brome mosaic virus supported organic dye arrays. A volume much smaller than the wavelength of light, high chromophore densities, and a near perfectly symmetric molecular arrangement, create a set of circumstances in which the photophysical properties of multichromophore particles are altered from those of individual chromophores. This discovery may lead to new viromimetic vectors with qualitatively enhanced optical emission properties, complemented by possibilities of further chemical or genetic manipulation for targeting in biomedical applications, and potentially other uses that could benefit from bright, deep-subwavelength light sources.

\section{Methods}

\subsection{BMV Purification}

wtBMV was expressed in \emph{Nicotiana benthamiana via Agrobacterium}-mediated gene delivery according to a previously described protocol\cite{Gopinath2007a}. Seven days after infection, \emph{N. benthamiana} leaves were homogenized in virus buffer [250 mM NaOAc, 10 mM \ce{MgCl2} (pH 4.5)] and then centrifuged at 6000 rpm for 25 minutes using an Eppendorf F-35-6-30 rotor. The supernatant was layered on a 10\% sucrose cushion in virus buffer and centrifuged at 26,000 rpm for 3 hours using a Beckman SW 32Ti rotor. The pellets were resuspended in 38.5\% CsCl (w/v, virus buffer) and were kept overnight in the cold room. Next day, solution was centrifuged at 10,000 rpm for 10 minutes in the BIO-Rad to remove the undissolved pellets. After that, supernatant was centrifuged at 45,000 rpm at 4$^\circ$C in a 100 Ti rotor. The resulting virus band was collected and dialyzed against SAMA (50 mM NaOAc, 8 mM \ce{Mg(OAc)2} (pH 4.6)) buffer for 24 hours, with three changes. Virus concentration was measured by UV-Vis spectrophotometry with a NanoDrop\texttrademark~instrument.

\subsection{Chemical Conjugation of wtBMV with Fluorescent Dyes}

Surface exposed lysines of wtBMV were labelled with Oregon Green 488\texttrademark carboxylic acid, succinimidyl ester. Concentrated stock of wtBMV was diluted in phosphate buffer (100 mM, pH 8.5) or sodium bicarbonate buffer (100 mM, pH 7.5) to a concentration of $10^{13}$ particles/ml. The dye was freshly prepared in DMSO per manufacturer instructions and mixed with the BMV solution at different ratios ranging from 1000 to 20000 dyes/capsid. The reaction was left for 2-3 hours at room temperature. Free dye was then removed via dialysis or filter wash with virus storage buffer at pH 4.5. The wash step was repeated until there was no presence of free dye in supernatant by UV-Vis.

\subsection{Chemical Conjugation via Flexible Linker}

The concentrated stock of wt-BMV was diluted in phosphate buffer (100 mM, pH 7) to a concentration of $10^{13}$ particles/ml, followed by reacting with 100$\times$ excess of Sulfo-SMCC (sulfosuccinimidyl 4-(N-maleimidomethyl) cyclohexane-1-carboxylate, ThermoFisher Scientific) for 1 hour, at room temperature. Excess SMCC was removed from the mixture by passing it through a Sephadex column (PD-10) with multiple washes using SAMA buffer with pH 6. The virus concentration  in the BMV-SMCC mixture was estimated via Nanodrop spectrometry, and then a bi-functionalized polyethylene glycol terminated with a thiol and an amine at the other end was mixed. The mixture was left to react for another hour at room temperature. Concentration of the BMV-SMCC linker was determined and the solution was characterized by mass spectrometry. Then, OG488 was added insimilar proportions as for BMV-OG samples, followed by removal of excess dye using dialysis.

\subsection{Preparation of Dye-coated Silica Nanoparticles}

Amine-modification of  $27\pm3$ nm silica nanoparticle (purchased from Sigma-Aldrich) was carried out through a post-synthesis grafting method\cite{Lehman2014} with (3-Aminopropyl)triethoxysilane (APTES, Sigma-Aldrich). After purification,  nanoparticles were functionalized with Oregon Green 488 carboxylic acid, succinimidyl ester following the same procedure as for BMV. For labelling efficiency calculations, the particle concentration was estimated from thermal gravimetric analysis (TGA), while dye concentration was estimated from UV-Vis absorbance. Sample morphology and particle size distribution were characterized by TEM.

\subsection{Transmission Electron Microscopy}

Electron-transparent samples were prepared by placing 10 $\mu$L of dilute sample onto a carbon-coated copper grid. After 10 min, the excess solution on the grid was removed with filter paper. The sample was stained with 10 $\mu$L of 2\% uranyl acetate for 10 minutes, and the excess solution was removed by blotting with filter paper. The sample was then left to dry for several minutes. Images were acquired at an accelerating voltage of 80 kV on a JEOL JEM1010 transmission electron microscope, and analyzed with the ImageJ Processing Toolkit to estimate particle diameters and for overall morphological characterization.

\subsection{Cryo-Electron Microscopy}

Cryo-EM frozen hydrated specimen was prepared as follows: 4 $\mu$L sample of Oregon Green labelled BMV capsid was loaded onto a glow-discharged Quantifoil R2/2 300 meshed copper grid coated with an ultrathin layer of carbon film. The grid was quickly frozen by using a FEI Vitrobot\texttrademark~Mark III. The humidity in the chamber was set to 100\% and the blotting time was 4 sec. The frozen hydrated specimen was then transferred to a Gatan 626 cryo-holder and kept under liquid nitrogen temperature (approximately -172$^\circ$C) for subsequent data collection. The grid was imaged using a 300-kV JEOL JEM-3200FS equipped with in-column energy filter and a DE-64 direct electron detection camera. The energy slit width was set to 20 eV and a nominal magnification of $\times$60000 (equivalent to 0.854 \AA~per pixel) was used to collect the 141, 25-frame movie stacks. The movie stacks were aligned, dose-weighted, and summed using MotionCor2\cite{Zheng2017a}. The first 2 and the last 10 frames were discarded, leaving the total dose to be less than 30 e$^-$/\AA$^2$. The defocus value for each micrograph was estimated using the CTFFIND4 program\cite{Rohou2015a}. A total of 31378 particles were semi-manually picked using e2boxer.py\cite{Tang2007a}. Reference-free 2D classification was performed on the whole particle set using Relion 2.1\cite{Scheres2012a} and the particles within class averages that are shown to be either clear noise or significant overlap were discarded at this point. The class averages that showed clear capsomer density were then used to build the initial low-resolution model using e2initialmodel.py. This initial model was further refined using Relion 3D auto-refine with a total of 12522 particles and a post-processing procedure to a final resolution of 7 \AA. The 3D model of Oregon Green labeled BMV was visualized and analyzed using UCSF Chimera\cite{Pettersen2004a}. The atomic coordinates of native BMV (PDB code 1SJ9) solved by X-ray crystallography was fitted into the cryo-EM density to analyze the potential binding positions for Oregon Green dyes.

\subsection{Steady-state Absorption and Fluorescence Emission Spectroscopy}

Absorbance spectra were recorded with a Varian Cary 100 Bio instrument. The concentration of wtBMV and the labeling efficiency of dyes were obtained from the UV-Visible absorbance spectra.  Fluorescence measurements were done on a QuantaMaster\texttrademark~fluorescence spectrometer (Horiba) with the following parameters: excitation wavelength: 488 nm; emission wavelength: 515 nm. The excitation and steady state emission spectra of 100 nM solutions were reported as averages from at least three independent measurements. wt-BMV and OG488 were used as reference standards whose optical characteristics were compared against labeled-BMV.

\subsection{Fluorescence Lifetime Imaging Microscopy}

Single particle fluorescence lifetime measurements were carried out with a Leica TCS SP8 SMD instrument in FLIM configuration. The sample, in the form of a dispersion of particles adsorbed on a silica glass coverslip in water, was excited at a wavelength of 488 nm from a filtered (10 nm bandwidth) supercontinuum source and pulse duration of $\sim120$ ps at 40 MHz. The maximum time-averaged power was $500~\mu$W at the sample. The collection spectral window was 500 nm to 550 nm. Data analysis was done with SymPhoTime 64\texttrademark software (PicoQuant, GmbH), the ImageJ Processing Toolkit (NIH), and with IgorPro (Wavemetrics, Inc.).

\subsection{Matrix Assisted Laser Desorption/Ionization--Time of Flight Spectrometry}

Fluorescent dye -- labeled BMV samples were buffer-exchanged to water. The diluted samples were mixed with the $\alpha$-cyano-4-hydroxycinnamic acid (CCA) matrix at a 1:5 ratio (varies depending upon the concentration of the sample). Myoglobin and CytochromeC protein mixture was used as the calibrant, which was introduced in 1:1:5 ratio with respect to the CCA matrix. MALDI-TOF mass spectrometry was done on a Bruker Autoflex III in the linear mode. For laser desorption, a 355 nm frequency tripled beam Nd-YAG laser was incident on the 384-well steel plate of $\sim 1 \mu$L spot size.

\subsection{Pulsed Excitation Fluorescence Spectroscopy}

An optical parametric amplifier (OPA, Coherent Inc.) pumped by a regenerative amplifier at 250 kHz repetition rate, and 170 fs pulsewidth, and having the output tuned to $\lambda = 490$ nm ($<P> = 100~\mu$W average power at the sample) was used as an excitation source. The 490 nm excitation beam was focused in the aqueous solution sample with a fused silica lens of 5 cm focal distance. Fluorescence from the sample was collected and spectrally analyzed using an Acton SpectraPro-300i monochromator fit with a TE-cooled CCD camera (iDus, Andor). Andor Solis software was used for data collection and IgorPro software was used for data processing and analysis.

\bibliography{IrinaBMVCollRefs}

\providecommand{\latin}[1]{#1}
\providecommand*\mcitethebibliography{\thebibliography}
\csname @ifundefined\endcsname{endmcitethebibliography}
  {\let\endmcitethebibliography\endthebibliography}{}
\begin{mcitethebibliography}{50}
\providecommand*\natexlab[1]{#1}
\providecommand*\mciteSetBstSublistMode[1]{}
\providecommand*\mciteSetBstMaxWidthForm[2]{}
\providecommand*\mciteBstWouldAddEndPuncttrue
  {\def\EndOfBibitem{\unskip.}}
\providecommand*\mciteBstWouldAddEndPunctfalse
  {\let\EndOfBibitem\relax}
\providecommand*\mciteSetBstMidEndSepPunct[3]{}
\providecommand*\mciteSetBstSublistLabelBeginEnd[3]{}
\providecommand*\EndOfBibitem{}
\mciteSetBstSublistMode{f}
\mciteSetBstMaxWidthForm{subitem}{(\alph{mcitesubitemcount})}
\mciteSetBstSublistLabelBeginEnd
  {\mcitemaxwidthsubitemform\space}
  {\relax}
  {\relax}

\bibitem[Tipirneni \latin{et~al.}(2017)Tipirneni, Warram, Moore, Prince,
  de~Boer, Jani, Wapnir, Liao, Bouvet, Behnke, Hawn, Poultsides, Vahrmeijer,
  Carroll, Zinn, and Rosenthal]{Tipirneni2017}
Tipirneni,~K.~E.; Warram,~J.~M.; Moore,~L.~S.; Prince,~A.~C.; de~Boer,~E.;
  Jani,~A.~H.; Wapnir,~I.~L.; Liao,~J.~C.; Bouvet,~M.; Behnke,~N.~K.
  \latin{et~al.}  {Oncologic Procedures Amenable to Fluorescence-guided
  Surgery}. \emph{Ann. Surg.} \textbf{2017}, \emph{266}, 36--47\relax
\mciteBstWouldAddEndPuncttrue
\mciteSetBstMidEndSepPunct{\mcitedefaultmidpunct}
{\mcitedefaultendpunct}{\mcitedefaultseppunct}\relax
\EndOfBibitem
\bibitem[Song \latin{et~al.}(2018)Song, Wang, Zhong, Chu, Su, and He]{Song2018}
Song,~B.; Wang,~H.; Zhong,~Y.; Chu,~B.; Su,~Y.; He,~Y. {Fluorescent and
  magnetic anti-counterfeiting realized by biocompatible multifunctional
  silicon nanoshuttle-based security ink}. \emph{Nanoscale} \textbf{2018},
  \emph{10}, 1617--1621\relax
\mciteBstWouldAddEndPuncttrue
\mciteSetBstMidEndSepPunct{\mcitedefaultmidpunct}
{\mcitedefaultendpunct}{\mcitedefaultseppunct}\relax
\EndOfBibitem
\bibitem[Stepanenko \latin{et~al.}(2011)Stepanenko, Stepanenko, Shcherbakova,
  Kuznetsova, Turoverov, and Verkhusha]{Stepanenko2011}
Stepanenko,~O.~V.; Stepanenko,~O.~V.; Shcherbakova,~D.~M.; Kuznetsova,~I.~M.;
  Turoverov,~K.~K.; Verkhusha,~V.~V. {Modern fluorescent proteins: from
  chromophore formation to novel intracellular applications.}
  \emph{Biotechniques} \textbf{2011}, \emph{51}, 313--318\relax
\mciteBstWouldAddEndPuncttrue
\mciteSetBstMidEndSepPunct{\mcitedefaultmidpunct}
{\mcitedefaultendpunct}{\mcitedefaultseppunct}\relax
\EndOfBibitem
\bibitem[Tsien(2008)]{Tsien2008}
Tsien,~R.~Y. {Roger Y. Tsien - Nobel Lecture: Constructing and Exploiting the
  Fluorescent Protein Paintbox}. 2008; \url{https://www.nobelprize.org/}\relax
\mciteBstWouldAddEndPuncttrue
\mciteSetBstMidEndSepPunct{\mcitedefaultmidpunct}
{\mcitedefaultendpunct}{\mcitedefaultseppunct}\relax
\EndOfBibitem
\bibitem[Valeur and Berberan-Santos(2013)Valeur, and
  Berberan-Santos]{Valeur2013}
Valeur,~B.; Berberan-Santos,~M. r.~N. \emph{John Wiley Sons}, 2nd ed.;
  Principles and Applications; John Wiley {\&} Sons: New York, 2013\relax
\mciteBstWouldAddEndPuncttrue
\mciteSetBstMidEndSepPunct{\mcitedefaultmidpunct}
{\mcitedefaultendpunct}{\mcitedefaultseppunct}\relax
\EndOfBibitem
\bibitem[Khitrova and Gibbs(2007)Khitrova, and Gibbs]{Khitrova2007a}
Khitrova,~G.; Gibbs,~H.~M. {Collective radiance}. \emph{Nat. Phys.}
  \textbf{2007}, \emph{3}, 84--85\relax
\mciteBstWouldAddEndPuncttrue
\mciteSetBstMidEndSepPunct{\mcitedefaultmidpunct}
{\mcitedefaultendpunct}{\mcitedefaultseppunct}\relax
\EndOfBibitem
\bibitem[Dicke(1954)]{Dicke1954}
Dicke,~R.~H. {Coherence in Spontaneous Radiation Processes}. \emph{Phys. Rev.}
  \textbf{1954}, \emph{93}, 99--110\relax
\mciteBstWouldAddEndPuncttrue
\mciteSetBstMidEndSepPunct{\mcitedefaultmidpunct}
{\mcitedefaultendpunct}{\mcitedefaultseppunct}\relax
\EndOfBibitem
\bibitem[Ball(2011)]{Ball2011}
Ball,~P. {Physics of life: The dawn of quantum biology}. \emph{Nature}
  \textbf{2011}, \emph{474}, 272--274\relax
\mciteBstWouldAddEndPuncttrue
\mciteSetBstMidEndSepPunct{\mcitedefaultmidpunct}
{\mcitedefaultendpunct}{\mcitedefaultseppunct}\relax
\EndOfBibitem
\bibitem[Scholes \latin{et~al.}(2011)Scholes, Fleming, Olaya-Castro, and van
  Grondelle]{Scholes2011}
Scholes,~G.~D.; Fleming,~G.~R.; Olaya-Castro,~A.; van Grondelle,~R. {Lessons
  from nature about solar light harvesting}. \emph{Nat. Chem.} \textbf{2011},
  \emph{3}, 763--774\relax
\mciteBstWouldAddEndPuncttrue
\mciteSetBstMidEndSepPunct{\mcitedefaultmidpunct}
{\mcitedefaultendpunct}{\mcitedefaultseppunct}\relax
\EndOfBibitem
\bibitem[Monshouwer \latin{et~al.}(1997)Monshouwer, Abrahamsson, van Mourik,
  and van Grondelle]{Monshouwer1997}
Monshouwer,~R.; Abrahamsson,~M.; van Mourik,~F.; van Grondelle,~R.
  {Superradiance and Exciton Delocalization in Bacterial Photosynthetic
  Light-Harvesting Systems}. \emph{J. Phys. Chem. B} \textbf{1997}, \emph{101},
  7241--7248\relax
\mciteBstWouldAddEndPuncttrue
\mciteSetBstMidEndSepPunct{\mcitedefaultmidpunct}
{\mcitedefaultendpunct}{\mcitedefaultseppunct}\relax
\EndOfBibitem
\bibitem[Purcell(1946)]{Purcell1946}
Purcell,~E.~M. {Spontaneous emission probabilities at radio frequencies}. Proc.
  Am. Phys. Soc. 1946; p 681\relax
\mciteBstWouldAddEndPuncttrue
\mciteSetBstMidEndSepPunct{\mcitedefaultmidpunct}
{\mcitedefaultendpunct}{\mcitedefaultseppunct}\relax
\EndOfBibitem
\bibitem[Vrehen \latin{et~al.}(1980)Vrehen, Schuurmans, and Polder]{Vrehen1980}
Vrehen,~Q.; Schuurmans,~M.; Polder,~D. {Superfluorescence: macroscopic quantum
  fluctuations in the time domain}. \emph{Nature} \textbf{1980}, \emph{285},
  70--71\relax
\mciteBstWouldAddEndPuncttrue
\mciteSetBstMidEndSepPunct{\mcitedefaultmidpunct}
{\mcitedefaultendpunct}{\mcitedefaultseppunct}\relax
\EndOfBibitem
\bibitem[Gross and Haroche(1982)Gross, and Haroche]{Gross1982a}
Gross,~M.; Haroche,~S. {Superradiance: An essay on the theory of collective
  spontaneous emission}. \emph{Phys. Rep.} \textbf{1982}, \emph{93},
  301--396\relax
\mciteBstWouldAddEndPuncttrue
\mciteSetBstMidEndSepPunct{\mcitedefaultmidpunct}
{\mcitedefaultendpunct}{\mcitedefaultseppunct}\relax
\EndOfBibitem
\bibitem[Benedict(1996)]{Benedict1996}
Benedict,~M. G. M.~G. \emph{{Super-radiance : multiatomic coherent emmission}};
  Institute of Physics Pub, 1996; p 326\relax
\mciteBstWouldAddEndPuncttrue
\mciteSetBstMidEndSepPunct{\mcitedefaultmidpunct}
{\mcitedefaultendpunct}{\mcitedefaultseppunct}\relax
\EndOfBibitem
\bibitem[Cong \latin{et~al.}(2016)Cong, Zhang, Wang, Noe, Belyanin, and
  Kono]{Cong2016}
Cong,~K.; Zhang,~Q.; Wang,~Y.; Noe,~G.~T.; Belyanin,~A.; Kono,~J. {Dicke
  superradiance in solids [Invited]}. \emph{J. Opt. Soc. Am. B} \textbf{2016},
  \emph{33}, C80\relax
\mciteBstWouldAddEndPuncttrue
\mciteSetBstMidEndSepPunct{\mcitedefaultmidpunct}
{\mcitedefaultendpunct}{\mcitedefaultseppunct}\relax
\EndOfBibitem
\bibitem[{Mallawa Arachchi} \latin{et~al.}(2018){Mallawa Arachchi}, Premaratne,
  and Maini]{MallawaArachchi2018}
{Mallawa Arachchi},~S.; Premaratne,~M.; Maini,~P.~K. {Superradiant Cancer
  Hyperthermia using a Buckyball Assembly of Quantum Dot Emitters}. \emph{IEEE
  J. Sel. Top. Quantum Electron.} \textbf{2018}, 1--1\relax
\mciteBstWouldAddEndPuncttrue
\mciteSetBstMidEndSepPunct{\mcitedefaultmidpunct}
{\mcitedefaultendpunct}{\mcitedefaultseppunct}\relax
\EndOfBibitem
\bibitem[Wang and Guo(2016)Wang, and Guo]{Wang2016}
Wang,~Y.; Guo,~L. {Nanomaterial-Enabled Neural Stimulation}. \emph{Front.
  Neurosci.} \textbf{2016}, \emph{10}\relax
\mciteBstWouldAddEndPuncttrue
\mciteSetBstMidEndSepPunct{\mcitedefaultmidpunct}
{\mcitedefaultendpunct}{\mcitedefaultseppunct}\relax
\EndOfBibitem
\bibitem[Skribanowitz \latin{et~al.}(1973)Skribanowitz, Herman, MacGillivray,
  and Feld]{Skribanowitz1973a}
Skribanowitz,~N.; Herman,~I.~P.; MacGillivray,~J.~C.; Feld,~M.~S. {Observation
  of Dicke Superradiance in Optically Pumped HF Gas}. \emph{Phys. Rev. Lett.}
  \textbf{1973}, \emph{30}, 309--312\relax
\mciteBstWouldAddEndPuncttrue
\mciteSetBstMidEndSepPunct{\mcitedefaultmidpunct}
{\mcitedefaultendpunct}{\mcitedefaultseppunct}\relax
\EndOfBibitem
\bibitem[Malcuit \latin{et~al.}(1987)Malcuit, Maki, Simkin, and
  Boyd]{Malcuit1987}
Malcuit,~M.~S.; Maki,~J.~J.; Simkin,~D.~J.; Boyd,~R.~W. {Transition from
  superfluorescence to amplified spontaneous emission}. \emph{Phys. Rev. Lett.}
  \textbf{1987}, \emph{59}, 1189--1192\relax
\mciteBstWouldAddEndPuncttrue
\mciteSetBstMidEndSepPunct{\mcitedefaultmidpunct}
{\mcitedefaultendpunct}{\mcitedefaultseppunct}\relax
\EndOfBibitem
\bibitem[Scheibner \latin{et~al.}(2007)Scheibner, Schmidt, Worschech, Forchel,
  Bacher, Passow, and Hommel]{Scheibner2007}
Scheibner,~M.; Schmidt,~T.; Worschech,~L.; Forchel,~A.; Bacher,~G.; Passow,~T.;
  Hommel,~D. {Superradiance of quantum dots}. \emph{Nat. Phys.} \textbf{2007},
  \emph{3}, 106--110\relax
\mciteBstWouldAddEndPuncttrue
\mciteSetBstMidEndSepPunct{\mcitedefaultmidpunct}
{\mcitedefaultendpunct}{\mcitedefaultseppunct}\relax
\EndOfBibitem
\bibitem[Spano and Mukamel(1989)Spano, and Mukamel]{Spano1989}
Spano,~F.~C.; Mukamel,~S. {Superradiance in molecular aggregates}. \emph{J.
  Chem. Phys.} \textbf{1989}, \emph{91}, 683--700\relax
\mciteBstWouldAddEndPuncttrue
\mciteSetBstMidEndSepPunct{\mcitedefaultmidpunct}
{\mcitedefaultendpunct}{\mcitedefaultseppunct}\relax
\EndOfBibitem
\bibitem[Bradac \latin{et~al.}(2017)Bradac, Johnsson, van Breugel, Baragiola,
  Martin, Juan, Brennen, and Volz]{Bradac2017}
Bradac,~C.; Johnsson,~M.~T.; van Breugel,~M.; Baragiola,~B.~Q.; Martin,~R.;
  Juan,~M.~L.; Brennen,~G.~K.; Volz,~T. {Room-temperature spontaneous
  superradiance from single diamond nanocrystals}. \emph{Nat. Commun.}
  \textbf{2017}, \emph{8}, 1205\relax
\mciteBstWouldAddEndPuncttrue
\mciteSetBstMidEndSepPunct{\mcitedefaultmidpunct}
{\mcitedefaultendpunct}{\mcitedefaultseppunct}\relax
\EndOfBibitem
\bibitem[Lucas \latin{et~al.}(2002)Lucas, Larson, and McPherson]{Lucas2002}
Lucas,~R.~W.; Larson,~S.~B.; McPherson,~A. {The crystallographic structure of
  brome mosaic virus.} \emph{J. Mol. Biol.} \textbf{2002}, \emph{317},
  95--108\relax
\mciteBstWouldAddEndPuncttrue
\mciteSetBstMidEndSepPunct{\mcitedefaultmidpunct}
{\mcitedefaultendpunct}{\mcitedefaultseppunct}\relax
\EndOfBibitem
\bibitem[Spano(2010)]{Spano2010}
Spano,~F.~C. {The Spectral Signatures of Frenkel Polarons in H- and
  J-Aggregates}. \emph{Acc. Chem. Res.} \textbf{2010}, \emph{43},
  429--439\relax
\mciteBstWouldAddEndPuncttrue
\mciteSetBstMidEndSepPunct{\mcitedefaultmidpunct}
{\mcitedefaultendpunct}{\mcitedefaultseppunct}\relax
\EndOfBibitem
\bibitem[Wen and Steinmetz(2016)Wen, and Steinmetz]{Wen2016}
Wen,~A.~M.; Steinmetz,~N.~F. {Design of virus-based nanomaterials for medicine,
  biotechnology, and energy}. \emph{Chem. Soc. Rev.} \textbf{2016}, \emph{45},
  4074--4126\relax
\mciteBstWouldAddEndPuncttrue
\mciteSetBstMidEndSepPunct{\mcitedefaultmidpunct}
{\mcitedefaultendpunct}{\mcitedefaultseppunct}\relax
\EndOfBibitem
\bibitem[Soto \latin{et~al.}(2006)Soto, Blum, Vora, Lebedev, Meador, Won,
  Chatterji, Johnson, and Ratna]{Soto2006}
Soto,~C.~M.; Blum,~A.~S.; Vora,~G.~J.; Lebedev,~N.; Meador,~C.~E.; Won,~A.~P.;
  Chatterji,~A.; Johnson,~J.~E.; Ratna,~B.~R. {Fluorescent Signal Amplification
  of Carbocyanine Dyes Using Engineered Viral Nanoparticles}. \emph{J. Am.
  Chem. Soc.} \textbf{2006}, \emph{128}, 5184--5189\relax
\mciteBstWouldAddEndPuncttrue
\mciteSetBstMidEndSepPunct{\mcitedefaultmidpunct}
{\mcitedefaultendpunct}{\mcitedefaultseppunct}\relax
\EndOfBibitem
\bibitem[Allen and Peters(1970)Allen, and Peters]{Allen1970}
Allen,~L.; Peters,~G. {Superradiance, coherence brightening and amplified
  spontaneous emission}. \emph{Phys. Lett. A} \textbf{1970}, \emph{31},
  95--96\relax
\mciteBstWouldAddEndPuncttrue
\mciteSetBstMidEndSepPunct{\mcitedefaultmidpunct}
{\mcitedefaultendpunct}{\mcitedefaultseppunct}\relax
\EndOfBibitem
\bibitem[Shammah \latin{et~al.}(2017)Shammah, Lambert, Nori, and {De
  Liberato}]{Shammah2017a}
Shammah,~N.; Lambert,~N.; Nori,~F.; {De Liberato},~S. {Superradiance with local
  phase-breaking effects}. \emph{Phys. Rev. A} \textbf{2017}, \emph{96},
  023863\relax
\mciteBstWouldAddEndPuncttrue
\mciteSetBstMidEndSepPunct{\mcitedefaultmidpunct}
{\mcitedefaultendpunct}{\mcitedefaultseppunct}\relax
\EndOfBibitem
\bibitem[Shahbazyan(2017)]{Shahbazyan2017}
Shahbazyan,~T.~V. {Mode Volume, Energy Transfer, and Spaser Threshold in
  Plasmonic Systems with Gain}. \emph{ACS Photonics} \textbf{2017}, \emph{4},
  1003--1008\relax
\mciteBstWouldAddEndPuncttrue
\mciteSetBstMidEndSepPunct{\mcitedefaultmidpunct}
{\mcitedefaultendpunct}{\mcitedefaultseppunct}\relax
\EndOfBibitem
\bibitem[Samuel \latin{et~al.}(2009)Samuel, Namdas, and Turnbull]{Samuel2009}
Samuel,~I. D.~W.; Namdas,~E.~B.; Turnbull,~G.~A. {How to recognize lasing}.
  \emph{Nat. Photonics} \textbf{2009}, \emph{3}, 546--549\relax
\mciteBstWouldAddEndPuncttrue
\mciteSetBstMidEndSepPunct{\mcitedefaultmidpunct}
{\mcitedefaultendpunct}{\mcitedefaultseppunct}\relax
\EndOfBibitem
\bibitem[Allen and Peters(1973)Allen, and Peters]{Allen1973}
Allen,~L.; Peters,~G.~I. {Amplified Spontaneous Emission and External Signal
  Amplification in an Inverted Medium}. \emph{Phys. Rev. A} \textbf{1973},
  \emph{8}, 2031--2047\relax
\mciteBstWouldAddEndPuncttrue
\mciteSetBstMidEndSepPunct{\mcitedefaultmidpunct}
{\mcitedefaultendpunct}{\mcitedefaultseppunct}\relax
\EndOfBibitem
\bibitem[Gelbart and Knobler(2009)Gelbart, and Knobler]{Gelbart2009}
Gelbart,~W.~M.; Knobler,~C.~M. {VIROLOGY: Pressurized Viruses}. \emph{Science}
  \textbf{2009}, \emph{323}, 1682--1683\relax
\mciteBstWouldAddEndPuncttrue
\mciteSetBstMidEndSepPunct{\mcitedefaultmidpunct}
{\mcitedefaultendpunct}{\mcitedefaultseppunct}\relax
\EndOfBibitem
\bibitem[Zeng(2017)]{Zeng2017b}
Zeng,~C. {Structure and mechanochemistry of icosahedral viruses and virus
  shells studied by atomic force microscopy}. Ph.D.\ thesis, Indiana
  University, Bloomington, 2017\relax
\mciteBstWouldAddEndPuncttrue
\mciteSetBstMidEndSepPunct{\mcitedefaultmidpunct}
{\mcitedefaultendpunct}{\mcitedefaultseppunct}\relax
\EndOfBibitem
\bibitem[Arakawa and Timasheff(1985)Arakawa, and Timasheff]{Arakawa1985}
Arakawa,~T.; Timasheff,~S.~N. {Mechanism of polyethylene glycol interaction
  with proteins}. \emph{Biochemistry} \textbf{1985}, \emph{24},
  6756--6762\relax
\mciteBstWouldAddEndPuncttrue
\mciteSetBstMidEndSepPunct{\mcitedefaultmidpunct}
{\mcitedefaultendpunct}{\mcitedefaultseppunct}\relax
\EndOfBibitem
\bibitem[Knowles \latin{et~al.}(2015)Knowles, Shkel, Phan, Sternke, Lingeman,
  Cheng, Cheng, O'Connor, and Record]{Knowles2015a}
Knowles,~D.~B.; Shkel,~I.~A.; Phan,~N.~M.; Sternke,~M.; Lingeman,~E.;
  Cheng,~X.; Cheng,~L.; O'Connor,~K.; Record,~M.~T. {Chemical Interactions of
  Polyethylene Glycols (PEGs) and Glycerol with Protein Functional Groups:
  Applications to Effects of PEG and Glycerol on Protein Processes}.
  \emph{Biochemistry} \textbf{2015}, \emph{54}, 3528--3542\relax
\mciteBstWouldAddEndPuncttrue
\mciteSetBstMidEndSepPunct{\mcitedefaultmidpunct}
{\mcitedefaultendpunct}{\mcitedefaultseppunct}\relax
\EndOfBibitem
\bibitem[Svidzinsky \latin{et~al.}(2008)Svidzinsky, Chang, Lipkin, and
  Scully]{Svidzinsky2008}
Svidzinsky,~A.; Chang,~J.-T.; Lipkin,~H.; Scully,~M. {Fermi's golden rule does
  not adequately describe Dicke's superradiance}. \emph{J. Mod. Opt.}
  \textbf{2008}, \emph{55}, 3369--3378\relax
\mciteBstWouldAddEndPuncttrue
\mciteSetBstMidEndSepPunct{\mcitedefaultmidpunct}
{\mcitedefaultendpunct}{\mcitedefaultseppunct}\relax
\EndOfBibitem
\bibitem[Noriega \latin{et~al.}(2015)Noriega, Finley, Haberstroh, Geissler,
  Francis, and Ginsberg]{Noriega2015}
Noriega,~R.; Finley,~D.~T.; Haberstroh,~J.; Geissler,~P.~L.; Francis,~M.~B.;
  Ginsberg,~N.~S. {Manipulating Excited-State Dynamics of Individual
  Light-Harvesting Chromophores through Restricted Motions in a Hydrated
  Nanoscale Protein Cavity.} \emph{J. Phys. Chem. B} \textbf{2015}, \emph{119},
  6963--73\relax
\mciteBstWouldAddEndPuncttrue
\mciteSetBstMidEndSepPunct{\mcitedefaultmidpunct}
{\mcitedefaultendpunct}{\mcitedefaultseppunct}\relax
\EndOfBibitem
\bibitem[Dietrich \latin{et~al.}(2016)Dietrich, Steude, Tropf, Schubert,
  Kronenberg, Ostermann, H{\"{o}}fling, and Gather]{Dietrich2016}
Dietrich,~C.~P.; Steude,~A.; Tropf,~L.; Schubert,~M.; Kronenberg,~N.~M.;
  Ostermann,~K.; H{\"{o}}fling,~S.; Gather,~M.~C. {An exciton-polariton laser
  based on biologically produced fluorescent protein}. \emph{Sci. Adv.}
  \textbf{2016}, \emph{2}, e1600666\relax
\mciteBstWouldAddEndPuncttrue
\mciteSetBstMidEndSepPunct{\mcitedefaultmidpunct}
{\mcitedefaultendpunct}{\mcitedefaultseppunct}\relax
\EndOfBibitem
\bibitem[Paul(1982)]{Paul1982}
Paul,~H. {Photon antibunching}. \emph{Rev. Mod. Phys.} \textbf{1982},
  \emph{54}, 1061--1102\relax
\mciteBstWouldAddEndPuncttrue
\mciteSetBstMidEndSepPunct{\mcitedefaultmidpunct}
{\mcitedefaultendpunct}{\mcitedefaultseppunct}\relax
\EndOfBibitem
\bibitem[Sun \latin{et~al.}(2007)Sun, DuFort, Daniel, Murali, Chen, Gopinath,
  Stein, De, Rotello, Holzenburg, Kao, and Dragnea]{Sun2007}
Sun,~J.; DuFort,~C.; Daniel,~M.-C.; Murali,~A.; Chen,~C.; Gopinath,~K.;
  Stein,~B.; De,~M.; Rotello,~V.~M.; Holzenburg,~A. \latin{et~al.}
  {Core-controlled polymorphism in virus-like particles}. \emph{Proc. Natl.
  Acad. Sci. U. S. A.} \textbf{2007}, \emph{104}, 1354--1359\relax
\mciteBstWouldAddEndPuncttrue
\mciteSetBstMidEndSepPunct{\mcitedefaultmidpunct}
{\mcitedefaultendpunct}{\mcitedefaultseppunct}\relax
\EndOfBibitem
\bibitem[Brillault \latin{et~al.}(2017)Brillault, Jutras, Dashti, Thuenemann,
  Morgan, Lomonossoff, Landsberg, and Sainsbury]{Brillault2017}
Brillault,~L.; Jutras,~P.~V.; Dashti,~N.; Thuenemann,~E.~C.; Morgan,~G.;
  Lomonossoff,~G.~P.; Landsberg,~M.~J.; Sainsbury,~F. {Engineering Recombinant
  Virus-like Nanoparticles from Plants for Cellular Delivery}. \emph{ACS Nano}
  \textbf{2017}, \emph{11}, 3476--3484\relax
\mciteBstWouldAddEndPuncttrue
\mciteSetBstMidEndSepPunct{\mcitedefaultmidpunct}
{\mcitedefaultendpunct}{\mcitedefaultseppunct}\relax
\EndOfBibitem
\bibitem[Lang and Chin(2014)Lang, and Chin]{Lang2014}
Lang,~K.; Chin,~J.~W. {Cellular Incorporation of Unnatural Amino Acids and
  Bioorthogonal Labeling of Proteins}. \emph{Chem. Rev.} \textbf{2014},
  \emph{114}, 4764--4806\relax
\mciteBstWouldAddEndPuncttrue
\mciteSetBstMidEndSepPunct{\mcitedefaultmidpunct}
{\mcitedefaultendpunct}{\mcitedefaultseppunct}\relax
\EndOfBibitem
\bibitem[Gopinath and Kao(2007)Gopinath, and Kao]{Gopinath2007a}
Gopinath,~K.; Kao,~C.~C. {Replication-independent long-distance trafficking by
  viral RNAs in Nicotiana benthamiana.} \emph{Plant Cell} \textbf{2007},
  \emph{19}, 1179--91\relax
\mciteBstWouldAddEndPuncttrue
\mciteSetBstMidEndSepPunct{\mcitedefaultmidpunct}
{\mcitedefaultendpunct}{\mcitedefaultseppunct}\relax
\EndOfBibitem
\bibitem[Lehman \latin{et~al.}(2014)Lehman, Tataurova, Mueller, Mariappan, and
  Larsen]{Lehman2014}
Lehman,~S.~E.; Tataurova,~Y.; Mueller,~P.~S.; Mariappan,~S. V.~S.;
  Larsen,~S.~C. {Ligand Characterization of Covalently Functionalized
  Mesoporous Silica Nanoparticles: An NMR Toolbox Approach}. \emph{J. Phys.
  Chem. C} \textbf{2014}, \emph{118}, 29943--29951\relax
\mciteBstWouldAddEndPuncttrue
\mciteSetBstMidEndSepPunct{\mcitedefaultmidpunct}
{\mcitedefaultendpunct}{\mcitedefaultseppunct}\relax
\EndOfBibitem
\bibitem[Zheng \latin{et~al.}(2017)Zheng, Palovcak, Armache, Verba, Cheng, and
  Agard]{Zheng2017a}
Zheng,~S.~Q.; Palovcak,~E.; Armache,~J.-P.; Verba,~K.~A.; Cheng,~Y.;
  Agard,~D.~A. {MotionCor2: anisotropic correction of beam-induced motion for
  improved cryo-electron microscopy}. \emph{Nat. Methods} \textbf{2017},
  \emph{14}, 331--332\relax
\mciteBstWouldAddEndPuncttrue
\mciteSetBstMidEndSepPunct{\mcitedefaultmidpunct}
{\mcitedefaultendpunct}{\mcitedefaultseppunct}\relax
\EndOfBibitem
\bibitem[Rohou and Grigorieff(2015)Rohou, and Grigorieff]{Rohou2015a}
Rohou,~A.; Grigorieff,~N. {CTFFIND4: Fast and accurate defocus estimation from
  electron micrographs}. \emph{J. Struct. Biol.} \textbf{2015}, \emph{192},
  216--221\relax
\mciteBstWouldAddEndPuncttrue
\mciteSetBstMidEndSepPunct{\mcitedefaultmidpunct}
{\mcitedefaultendpunct}{\mcitedefaultseppunct}\relax
\EndOfBibitem
\bibitem[Tang \latin{et~al.}(2007)Tang, Peng, Baldwin, Mann, Jiang, Rees, and
  Ludtke]{Tang2007a}
Tang,~G.; Peng,~L.; Baldwin,~P.~R.; Mann,~D.~S.; Jiang,~W.; Rees,~I.;
  Ludtke,~S.~J. {EMAN2: An extensible image processing suite for electron
  microscopy}. \emph{J. Struct. Biol.} \textbf{2007}, \emph{157}, 38--46\relax
\mciteBstWouldAddEndPuncttrue
\mciteSetBstMidEndSepPunct{\mcitedefaultmidpunct}
{\mcitedefaultendpunct}{\mcitedefaultseppunct}\relax
\EndOfBibitem
\bibitem[Scheres(2012)]{Scheres2012a}
Scheres,~S.~H. {RELION: Implementation of a Bayesian approach to cryo-EM
  structure determination}. \emph{J. Struct. Biol.} \textbf{2012}, \emph{180},
  519--530\relax
\mciteBstWouldAddEndPuncttrue
\mciteSetBstMidEndSepPunct{\mcitedefaultmidpunct}
{\mcitedefaultendpunct}{\mcitedefaultseppunct}\relax
\EndOfBibitem
\bibitem[Pettersen \latin{et~al.}(2004)Pettersen, Goddard, Huang, Couch,
  Greenblatt, Meng, and Ferrin]{Pettersen2004a}
Pettersen,~E.~F.; Goddard,~T.~D.; Huang,~C.~C.; Couch,~G.~S.;
  Greenblatt,~D.~M.; Meng,~E.~C.; Ferrin,~T.~E. {UCSF Chimera - A visualization
  system for exploratory research and analysis}. \emph{J. Comput. Chem.}
  \textbf{2004}, \emph{25}, 1605--1612\relax
\mciteBstWouldAddEndPuncttrue
\mciteSetBstMidEndSepPunct{\mcitedefaultmidpunct}
{\mcitedefaultendpunct}{\mcitedefaultseppunct}\relax
\EndOfBibitem
\end{mcitethebibliography}

\begin{acknowledgement}

The work was supported by the Army Research Office, under award W911NF-17-1-0329, and by the National Science Foundation, under award CBET 1803440 and 1808027, and by the IU Nanocharacterization Facility. 

\end{acknowledgement}

\begin{suppinfo}

Contains details on experimental procedures, and supplemental figures supporting the main text.

\end{suppinfo}

\end{document}